\documentclass[10point,a4paper]{article}
\usepackage{graphicx}
\usepackage{amsmath,amssymb}
\usepackage{array,blindtext}

\newcommand\beq{\begin{equation}}
\newcommand\eeq{\end{equation}}
\newcommand\bea{\begin{eqnarray}}
\newcommand\eea{\end{eqnarray}}

\begin{document}
\title{Exact Simulation of Loops in Lattice Gauge Theory}
\author{Indrakshi Raychowdhury\footnote{indrakshi@imsc.res.in}\\
The Institute of Mathematical Sciences, \\\vspace{0.7cm} CIT-Campus, Taramani,
Chennai, India}
\maketitle
\begin{abstract}
We exploit the local loop dynamics calculated in prepotential formulation to compute the pertrubation expansion in the strong coupling limit of lattice gauge theory. A new exact simulation technique is developed to simulate all possible loop states on an infinite lattice using exact BFACF algorithm. This loop perturbation numerical calculations are free from any limitation on lattice size. Using this, we calculate the correction of  string tension for SU(2) theory in 2+1 dimension upto order $g^{64}$.
\end{abstract}
\section{Introduction}

Lattice gauge theory, from its beginning has been extremely popular and advantageous primarily because of the plausibility of practical computations of high energy physics which was almost impossible to address in continuum. Just after the advent of lattice gauge theory in its Euclidean version \cite{wil}, a Hamiltonian version was also proposed in \cite{kogut}. However in the course of next three and a half decades, the later remains much unexplored primarily because of its not so well adaptation in numerical simulations, at least upto the level in which the first one has achieved. But, both the Hamiltonian and Euclidean path integral approach of lattice gauge theory has their own benefits over the other. The Hamiltonian formulation allows one to directly work with the fock space, vacuum structure and the spectrum of the theory which is not such straight forward in the other way. 

Lattice gauge theories do suffer from a huge redundant gauge degrees of freedom in general which is a primary obstacle in terms of computation cost. It is desirable to work within the physical degrees of freedom of the theory which is gauge independent or invariant. One can always achieve that either by making a suitable gauge choice or by formulating the theory completely in terms of gauge invariant variables such as Wilson loops and strings. The later prescription, although attempted by many theorists in the past \cite{loops}, has not been of much practical advantage as the loops formulation itself suffers from severe non-locality and proliferation of loop and string space.

There has been a recent approach in the Hamiltonian lattice gauge theory in terms of local loop degrees of freedom \cite{mm,pp_IR,prd}. This formulation is named as prepotential formulation, where prepotentials are basically harmonic oscillators belonging to the fundamental representation of the gauge group and are defined at each lattice site. These oscillators act as the fundamental operators of the theory in terms of which the original canonical variables are reconstructed. By definition, prepotentials are located at each lattice site. The most important feature of this formulation is that, in terms of prepotentials, the non-abelian part of the gauge group becomes ultra local, confined to each lattice site resulting local loop space or gauge invariant Hilbert space of the theory situated at each site. However, the original non-local loops such as Wilson loops  and strings do exist in the new formulation and are constructed by abelian weaving of local loops along neighbouring lattice sites. Prepotential formulation enables one  to construct exact orthonormal loop basis locally at each site of the lattice and compute the dynamics of the theory site by site.

It is worth mentioning that, there has been only a limited progress towards the numerical investigations in Hamiltonian lattice gauge theory since the advent of the subject compared to that in Euclidean version of the theory.  Most of the numerical studies within Hamiltonian framework till date uses the variational approach \cite{var_all}. However, the variational study primarily depend on the trial wave function which is not chosen systematically such that a systematic improvement in the understanding of the theory can be obtained. Some progresses have been done by applying many body techniques such as cluster expansion, specifically coupled cluster or linked cluster algorithms \cite{ccm} for lattice gauge theories. These cluster algorithms however involve abrupt truncation of the Hilbert space of the theory. Yet, they are advantageous in the sense that, they are in terms of gauge invariant variables or loops and hence free from any gauge redundancy. Note that, although classical Monte Carlo simulations provide a very powerful and accurate method for the
study of Euclidean lattice gauge theories, the quantum Monte Carlo simulations \cite{qmc} have been explored to a limited extend in the context of Hamiltonian lattice gauge theory. Moreover, these studies involve gauge field configurations likewise the classical Monte-Carlo for euclidean version of the theory and it is not possible to enforce Gauss's law explicitly in these calculations. Again, Monte-Carlo simulations in Hamiltonian framework, is not suitable to compute the string tensions and mass gaps directly as Hamiltonian eigenvalues, as one does in the strong-coupling representation. But in this approach, one is forced back to the more laborious approach used in Euclidean calculations by measuring the appropriate correlation
functions and estimating the mass gap as the inverse of the correlation length.

In this scenario, it is expected that, a better choice of strong-coupling basis, such as the `loop representation’
should avoid these problems and provide with a better numerical simulation scheme to explore Hamiltonian lattice gauge theory, which has not yet been demonstrated in literature so far. In this work, we proceed in this direction and develop a numerical simulation in terms of local orthogonal loop Hilbert space, whose dynamics is given in prepotential framework.
   
This work is the very first attempt to exploit the novel features of prepotential formulation in terms of explicit computations. This local loop formulation of lattice gauge theory do exist for any arbitrary SU(N) gauge group. But in this work we will concentrate on the simplest non-trivial case of $2+1$ dimensional non-abelian pure SU(2) gauge theory defined on lattice.  We will take a semi-analytic approach in doing so. We simulate all possible loop configurations of the theory, starting from the strong coupling ground state numerically using a new exact simulation technique developed by us and calculate the Rayleigh-Schr\"odinger perturbation expansion upto a high enough order (correcting upto $g^{-64}$) using the analytically calculated dynamics of the theory in prepotential formulation. This scheme of calculation is completely gauge invariant and is defined on a truly infinite spatial lattice. This saves the computation cost to a huge extent. Note that, the work presented in this paper is simulated and calculated on an ordinary PC within a few minutes.

The organization of the paper is as follows: We start with a brief outline of the loop perturbation expansion in Hamiltonian lattice gauge theory, which is Rayleigh Schr\"odinger perturbation expansion in section 2. In section 3, we introduce prepotentials and illustrate the Hamiltonian dynamics in terms of prepotential which is local. In section 4, we develop the algorithm for simulation of all loop configuration and illustrate how to construct the network of states, which is to be exploited in calculation of strong coupling perturbation expansion for the theory. We also illustrate the novelty of this algorithm by computing axial string tension semi-analytically using our algorithm. Next, in section 5, we discuss the results and finally conclude.

%

\section{Loop Perturbation Expansion in Hamiltonian Lattice Gauge Theory}


In this work we approach the Hamiltonian lattice gauge theory in its strong coupling limit in a complete gauge invariant way by numerical simulations of gauge invariant configurations of the theory. For the sake of completeness, we briefly discuss the Hamiltonian for the theory and the  perturbation corrections to its spectrum in strong coupling limit. Note that, in this work we will confine ourselves to the SU(2) gauge group and $2+1$ dimension for simplicity. In 2+1 dimensions, we consider discretization in the two spatial dimension only, keeping time continuous. The lattice is just a 2-dimensional square lattice.

The Lattice gauge theory Hamiltonian \cite{kogut} is basically a many body Hamiltonian constructed in terms of the canonical conjugate variables, i.e the left or right color electric field $E^{\mathrm a}_{L/R}(n,i)$ and link operator $U(n,i)$ located on each and every link $(n,i)$ originating at site $n$ along direction $i$ of the lattice, given by
\bea
\label{ham}
H=g^2 \sum_{\mbox{links}} E^{2}_{links}-\frac{1}{g^2}\sum_{\mbox{plaquettes}}\left(\mbox{Tr} U_{plaquette}+\mbox{Tr} U^\dagger_{plaquette}\right)
\eea
In (\ref{ham}),
$
U_{\mbox{plaquette}}=U(n,i)U(n+i,j)U^{\dagger}(n+j,i)U^{\dagger}(n,j)$ is product over links around a plaquette.
The canonical conjugate variables of the theory satisfy the commutation relation:
\bea
\left[E_L^{\mathrm a}(n,i),U^{\alpha}{}_{\beta}(n,i)\right] = - \left(\frac{\sigma^{\mathrm a}}{2}U(n,i)\right)^{\alpha}{}_{\beta},~~ 
~~~\left[E_R^{\mathrm a}(n+i,i),U^{\alpha}_{\beta}(n,i)\right] = \left(U(n,i) \frac{\sigma^{\mathrm a}}{2} \right)^{\alpha}{}_{\beta}. 
\label{ccr} 
\eea 
In (\ref{ccr}), 
$\mathrm a (=1,2,3)$ is the color index for SU(2) and $\frac{\sigma^{\mathrm a}}{2}$ are the generators in the fundamental representation of 
$SU(2)$ and satisfy: $[\frac{\sigma^{\mathrm a}}{2},\frac{\sigma^{\mathrm b}}{2}]=i\epsilon^{\mathrm{abc}}\frac{\sigma^{\mathrm c}}{2}$. The left and right electric fields follow SU(2) algebra:
\bea
\label{eec}
[E_L^{\mathrm a}(n,i),E_L^{\mathrm b}(n,i)] &=& i\epsilon_{\mathrm{abc}}E_L^{\mathrm c}(n,i),\nonumber \\
\left[E_R^{\mathrm a}(n,i),E_R^{\mathrm b}(n,i)\right] &=& i\epsilon_{\mathrm{abc}}E_R^{\mathrm c}(n,i),\\
\left[E_L^{\mathrm a}(n,i),E_R^{\mathrm b}(m,j)\right]&=& 0.\nonumber
\eea
Note that the right generators $E^\mathrm a_R(n+i,i)$ are the 
parallel transport of the left generator $E^\mathrm a_L(n,i)$ on the link $(n,i)$ as 
$ E_R(n+i,i)= - U^{\dagger}(n,i) E_L(n,i) U(n,i)
$, implying,
\bea
\label{E2=e2}
E^{2}_{links}\equiv \sum_{{\mathrm a}=1}^{3}E^{\mathrm a}(n,i)E^{\mathrm a}(n,i) =
\sum_{{\mathrm a}=1}^{3}E_L^{\mathrm a}(n,i)E_L^{\mathrm a}(n,i)
=\sum_{{\mathrm a}=1}^{3}E_R^{\mathrm a}(n+i,i)E_R^{\mathrm a}(n+i,i).
\eea
which appear in (\ref{ham}).

In the strong coupling limit of the theory, where $g^2\rightarrow \infty$, the Hamiltonian is dominated by the electric part, i.e
\bea
H_e=g^2 \sum_{\mbox{links}} E^{2}_{links}
\eea
The prime advantage of formulating gauge theories on lattice is that it is naturally confining at the strong coupling limit. One can perform the strong coupling perturbation expansion around this limit. In Hamiltonian formulation it is natural to carry out Rayleigh Schr\"odinger perturbation series expansion for lattice gauge theories with the unperturbed Hamiltonian
\bea
H_0\equiv H_e=g^2 \sum_{\mbox{links}} E^{2}_{links}
\eea
and interaction Hamiltonian given by
\bea
H_I\equiv H_{mag}= \frac{1}{g^2}\sum_{\mbox{plaquettes}}\left(\mbox{Tr} U_{plaquette}+\mbox{Tr} U^\dagger_{plaquette}\right).
\eea
At this point, we scale the Hamiltonian in the following way:
\bea
H\rightarrow g^2 H= \underbrace{\sum_{\mbox{links}} E^{2}_{links}}_{H_0}-\frac{1}{g^4}\underbrace{\sum_{\mbox{plaquettes}}\left(\mbox{Tr} U_{plaquette}+\mbox{Tr} U^\dagger_{plaquette}\right)}_{H_I} \equiv H_0- \frac{1}{y} H_I
\eea
The unperturbed Hamiltonian can be solved exactly, with the gauge invariant electric flux states or loops (or string) states as eigenstates and eigenvalues counting the net flux flowing around that loop.
This sector of the theory is completely confining showing finite mass gap and constant value of string tension.

Let us consider the unperturbed Hamiltonian and its spectrum first. Let $|n_i\rangle$ denote the $i^{th}$ excited eigenstate of $H_0$ with energy eigenvalue $E_0^i$, where the 0 index refers to unperturbed energy, i.e
\bea
H_0|n_i\rangle = E_0^{n_i}|n_i\rangle, ~~\mbox{for } i=0,1,2,3,.....
\eea
for $i=0$, we denote the strong coupling vacuum states by $|0\rangle$. The complete loop space is built by the successive  action of the interaction Hamiltonian on $|0\rangle$. The perturbation correction to the unperturbed spectrum is obtained from the full spectrum. For strong coupling lattice gauge theory, all odd order corrections do vanish and the net correction is obtained as
\bea
\label{Ecorr}
E_0=E_0^0+\frac{1}{y^2}E_0^{(2)}+\frac{1}{y^4}E_0^{(4)} +\ldots +\frac{1}{y^{2M}}E_0^{(2M)}+\ldots
\eea
 The even $(2M)$ order corrections are obtained by the expression:
\bea
\label{p2}
E_0^{(2M)}&=& \sum_{\{n_i\}\ne 0} \frac{\langle 0|H_I|n_1\rangle\langle n_1|H_I|n_2\rangle\ldots\langle n_{2M-1}|H_I|0\rangle  }{(E_0^0-E_0^{n_1}) (E_0^0-E_0^{n_2})\ldots (E_0^0-E_0^{n_{2M-1}})}\nonumber \\
&& - \sum_{p=1}^{M-1} E_0^{(2M-2p)} \sum_{\{n_i\}\ne 0}\Big[\frac{\langle 0|H_I|n_1\rangle\ldots\langle n_{p}|H_I|0\rangle  }{(E_0^0-E_0^{n_1})\ldots (E_0^0-E_0^{n_{p}})}\times \left( \sum_{i=1}^{p} \frac{1}{E_0^0-E_0^{n_i}} \right)\Big]
\eea
In the next section we discuss and evaluate this perturbation expansion in prepotential formulation, where we can calculate the dynamics of any arbitrary loop state locally at each site.

\section{Prepotentials and Hamiltonian Dynamics}

The recently developed prepotential formulation of Hamiltonian lattice gauge theory \cite{mm,pp_IR,prd} has introduced a new level of simplicity in the theory both in terms of theoretical understanding and practical computations. In this novel formulation, all the gauge invariant states or loop states has a local description forming an orthogonal and complete basis. Moreover, the dynamics is also local.

In this work, we will exploit the prepotential formulation to calculate  the strong coupling perturbation expansion for pure lattice gauge theory. We approach the problem semi-analytically, or to be more specific, our approach consists of two steps. First, we compute the local dynamics of loop states analytically and then numerically simulate all possible gauge invariant configurations and perform the sum of their contribution to the perturbation expansion numerically. The reason for going  beyond analytic treatment is the fact that the number of loop configurations diverge exponentially (which is clearly shown in our simulation) as one attempts to make higher and higher order corrections. Calculation of the degeneracy factor at each excited level is itself an analytically-unsolved problem in combinatorics. However we propose an algorithm, where all loop configurations are generated starting from strong coupling vacuum. In this particular work, for the purpose of illustration of the usefulness of the algorithm to compute physical quantities,  we consider the axial string tension to be calculated from  the string state connecting  a static quark and anti-quark pair as the physical loop configuration. 

Before entering into the dynamics of the specific system, it is worth having a brief review of the  prepotential formulation for SU(2) lattice gauge theory in $2+1$  dimension, which sets the platform of this particular work.

SU(2) prepotentials are basically two sets of Harmonic oscillator doublets $a^\dagger_\alpha(L/R), ~~\alpha=1,2$ associated to both the ends $(L/R)$ of all the links on the lattice. The canonical conjugate variables in the Kogut -Susskind Hamiltonian, namely, the color electric fields and link operators  on a link $(n,i)$ are rewritten in terms of the prepotentials in the following way, keeping all the commutation relations (\ref{ccr},\ref{eec}) invariant:
\bea
\label{sb1} 
\mbox{Left electric fields:} ~~ \quad \quad \quad E_L^{\mathrm a} &\equiv& 
a^{\dagger}(L)\frac{\sigma^{\mathrm a}}{2}a(L), \\ 
\label{sb2} 
\mbox{Right electric fields:} \quad ~ E_R^{\mathrm a} &\equiv & 
a^{\dagger}(R)\frac{\sigma^{\mathrm a}}{2}a(R). \nonumber \\
\label{sb3} 
\mbox{Link operator: ~~~} ~~U^{\alpha}{}_{\beta}&\equiv & \frac{1}{\sqrt{\hat n+1}}\left(\tilde{a}^{\dagger\alpha}(L) \, a^{\dagger}_\beta(R) 
+ a^\alpha(L)  \, \tilde{a}_\beta(R)\right)\frac{1}{\sqrt{\hat n+1}}
\eea
The above link operators are defined to satisfy the unitarity condition given by
\bea
UU^\dagger=U^\dagger U=1 ~~\& ~~ \mbox{Det}U
=1\eea
This novel prepotential approach has the following properties:
\begin{itemize}
\item Each link is subdivided into a left part and a right part. Even the link operator, in matrix form can also be written as \cite{pp_IR} $U=U_LU_R$.
\item The only connection between the left and right part of a link is via the abelian constraint
\bea 
\hat{n}(L) \equiv a^\dagger(L) \cdot a(L) = \hat{n}(R) \equiv a^\dagger(R) \cdot a(R) \equiv \hat{n} 
\label{noc} 
\eea 
derived from the condition $E_L^2=E_R^2$.
\end{itemize}
In summary, the prepotential formulation makes the non abelian part of the gauge group confined locally at each site and abelian fluxes flowing across neighbouring sites. As a result, this approach contains an ultra local description of loop states, operators and their dynamics as discussed in detail in \cite{mm,pp_IR,prd}. In context of this particular work, we highlight a few important observations regarding the dynamics of the theory as we are interested in the case of SU(2) gauge group and $2+1$ dimension. 

\begin{figure}[h]
\begin{center}
\includegraphics[width=0.2\textwidth,height=0.05\textwidth]{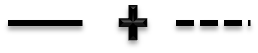}
\end{center}
\caption{Pictorial representation of link operator in prepotential formulation} 
\label{linkpp}
\end{figure}

\begin{figure}[h]
\begin{center}
\includegraphics[width=0.7\textwidth,height=0.2\textwidth]{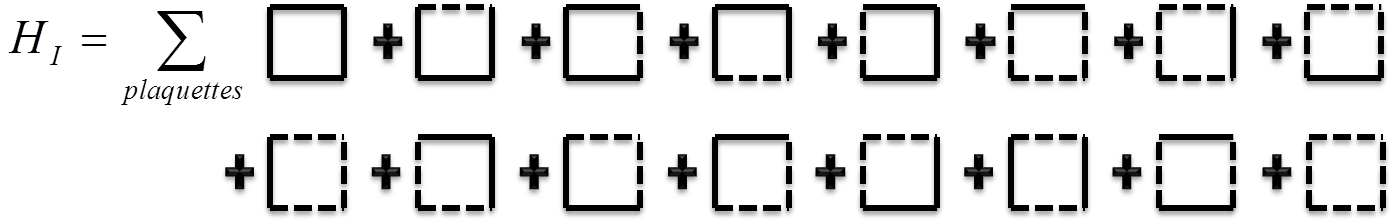}
\end{center}
\caption{The magnetic Hamiltonian for SU(2) lattice gauge theory} 
\label{h16}
\end{figure}

It is clear from the redefinition of link operator in terms of prepotentials in (\ref{sb3}), that each link operator is a sum of two terms. One of the terms contains only prepotential creation operators causing net increase the flux along the link by one unit and the other contains only prepotential annihilation operators which decrease the flux along that link by one unit, 
pictorially described in figure \ref{linkpp}.
The magnetic part of the Hamiltonian is defined around each plaquette consisting of four such link operators. Hence the full plaquette operator for a square lattice consists of $2^4=16$ terms as shown in figure \ref{h16}.

We now discuss the dynamics of loop states under this plaquette operator. The loop states in prepotential formulation are characterized by linking numbers at each site \cite{prd}. At each site of a two dimensional spatial lattice, there exist six distinct linking numbers  in order to characterize any arbitrary loop state together with three different constraints yielding three physical loop degree of freedom per site, as expected. The detailed dynamics of such orthonormal loops has been discussed in great detail in \cite{prd}. The dynamics around a plaquette can be realized more geometrically in terms of fusion variables as described in \cite{prd} which is again shown to be exactly equivalent to the same obtained in terms of original link variables. As we confine ourselves to gauge group SU(2) in this work, we do not enter into any of these complications but just quote the required dynamics discussed in \cite{mm}. 
The complete and orthonormal loop basis for this case is characterized by $3$ angular momentum quantum numbers per lattice site given by $|j_1,j_2,j_{12}\rangle$and hence, around a plaquette `abcd' as $$|j_{abcd}\rangle\equiv |j_1^a,j_2^a,j_{\bar 1}^a,j_{\bar 2}^a,j_{12}^a\rangle \times |j_1^b,j_2^b,j_{\bar 1}^b,j_{\bar 2}^b,j_{12}^b\rangle \times |j_1^c,j_2^c,j_{\bar 1}^c,j_{\bar 2}^c,j_{12}^c\rangle \times |j_1^d,j_2^d,j_{\bar 1}^d,j_{\bar 2}^d,j_{12}^d\rangle  $$
We can further identify
\bea
j_1^a=j_{\bar 1}^b\equiv j_1 ~& ~j_2^b=j_{\bar 2}^c\equiv j_2 \\
j_{\bar 1}^c=j_{ 1}^d\equiv j_{\bar 1}  ~&~ j_{\bar 2}^d=j_{2}^a\equiv j_{\bar 2} 
\eea 
as the four fluxes around a plaquette. 
The dynamics of such states under the plaquette action are obtained as \cite{mm}
\bea
\langle \bar{j}_{abcd}|{\textrm Tr}U_{abcd}| {j}_{abcd} \rangle  & = &  
{\cal M}_{abcd}  
{\left \{ \begin{array}{cccc}
{j}_{1} &  \bar{j}_{1} & \frac{1}{2}  \\
\bar{j}_{\bar 2} & {j}_{\bar 2}  & {j}^a_{12}   \\ 
\end{array} \right\}}   
{\left\{ \begin{array}{cccc}
{j}^b_{12} & \bar{j}^b_{12} & \frac{1}{2}  \\
\bar{j}_{1} & j_{1} & j_{\bar 2}^b\\
\end{array} \right \}}  
{\left\{ \begin{array}{cccc}
{j}^b_{12} & \bar{j}^b_{12} & \frac{1}{2}  \\
\bar{j}_{2} & j_{2} & j_1^b\\
\end{array} \right \}} 
\nonumber  \\ \nonumber  \\ &&
\hspace{1cm} {\left \{ \begin{array}{cccc}
{j}_{\bar 1} &  \bar{j}_{\bar 1} & \frac{1}{2}  \\
\bar{j}_{2} & {j}_{2}  & {j}^c_{12}   \\ 
\end{array} \right\}}   
{\left\{ \begin{array}{cccc}
{j}^d_{12} & \bar{j}^d_{12} & \frac{1}{2}  \\
\bar{j}_{\bar 1} & j_{\bar 1} & j_2^d\\
\end{array} \right \}}  
{\left\{ \begin{array}{cccc}
{j}^d_{12} & \bar{j}^d_{12} & \frac{1}{2}  \\
\bar{j}_{\bar 2} & j_{\bar 2} & j_{\bar 1}^d \\
\end{array}\right\}}. ~~~~
\label{dyna1} 
\eea
In (\ref{dyna1}), ${\cal M}_{abcd} \equiv D_{abcd} N_{abcd} P_{abcd}$ where: 
\bea 
&&D_{abcd} =  \delta_{j^a_{\bar 1},\bar{j}^a_{\bar 1}} 
\delta_{j^a_{\bar 2},\bar{j}^a_{\bar 2}} 
\delta_{j^a_{12},\bar{j}^a_{12}}  
\delta_{j_1^b,\bar{j}_1^b} \delta_{j_{\bar 2}^b,\bar{j}_{\bar 2}^b}  
\delta_{j^c_{1},\bar{j}^c_{1}} \delta_{j^c_{2},\bar{j}^c_{2}} \delta_{j^c_{12},\bar{j}^c_{12}} 
\delta_{j_2^d,\bar{j}_2^d} \delta_{j_{\bar 1}^d,\bar{j}_{\bar 1}^d}, 
\nonumber \\ 
\label{cnf} 
&& N_{abcd} =
{\Pi}\left(j_1,\bar{j}_{1}, j_2,\bar{j_2},j_3,\bar j_{\bar 1},j_{\bar 2},\bar j_{\bar 2},j^b_{12},\bar{j}^b_{12},
j^d_{12},\bar{j}^d_{12}\right) 
\\
&& P_{abcd} =  - (-1)^{j_1+j_2+j_1^b+j_{\bar 2}^b} (-1)^{j_{\bar 1}+j_{\bar 2}+j_{\bar 1}^d+j_2^d}  
\triangle(\bar{j}_1,\bar{j}_{\bar 2},j_{12}^{a}) 
\triangle(\bar{j}_2,\bar{j}_{\bar 1},j_{12}^{c}) \nonumber \\
&&~~~~~~~~~~~~~~~~~
\triangle(\bar{j}^b_{12},{j}^b_{12},\frac{1}{2})  
\triangle(\bar{j}^d_{12},{j}^d_{12},\frac{1}{2}).   
\nonumber 
\eea

In (\ref{cnf}), $D_{abcd}$ describes the trivial $\delta$ functions over the angular 
momenta which do not change under the action of the plaquette operator, 
$N_{abcd}$ and $P_{abcd}$ give the corresponding numerical and the phase factors respectively. 
The multiplicity factors are:  $\Pi(x,y,...) \equiv \sqrt{(2x+1)(2y+1)...}$ and 
$\triangle(x,y,z)$ represent the phase factors associated with a triangle with sides x, y, z: 
$\triangle(x,y,z)  \equiv  (-1)^{x+y+z} \Rightarrow \triangle(x,y,z) =\pm 1$.   

Having set the dynamics of Hamiltonian lattice gauge theory in terms of the prepotentials, let us now concentrate on the particular problem that we are interested in for this work. In order to calculate the string tension, our aim is to calculate the ground state energy of a string state connecting a static quark anti-quark pair as shown in figure \ref{qq}. 
\begin{figure}[h]
\label{qq}
\begin{center}
\includegraphics[width=0.3\textwidth,height=0.1\textwidth]{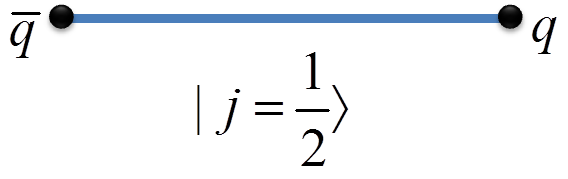}
\end{center}
\caption{A string state connecting a static $q,\bar q$ pair } 
\label{qq}
\end{figure}

To make a gauge invariant configuration, the flux string staring at $q$ and ending at $\bar q$ must carry one unit of flux, denoted by $|j=1/2\rangle$. The unperturbed energy eigenvalue for such a state is obtained in prepotential formulation as,
\bea
\sum_{\mbox{links}} E^{2}_{links}\Bigg[|j=1/2\rangle_{\mbox{along the string}}\Bigg]= \frac{3}{4}\times \mbox{length of the string in lattice units}
\eea
Hence, the lowest energy configuration of such a string state is the string connecting the quark-antiquark pair of minimum length, i.e along the shortest distance of the two points where the static sources are residing. The interaction Hamiltonian will act on such a string state and will deform it to another string state, which is again eigenstate of the unperturbed Hamiltonian with certain eigen value depending upon its length. 

\begin{figure}[ht]
\begin{minipage}[b]{.4\textwidth}
\centering
\includegraphics[width=1\textwidth]{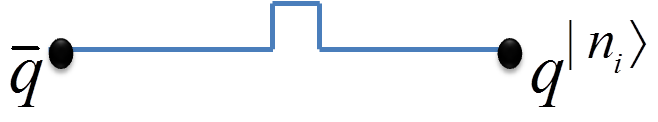}
\caption{An arbitrary string state}
\label{p1}
\end{minipage}
\hfill
\begin{minipage}[b]{.5\textwidth}
\centering
\vspace{0.1cm}
\includegraphics[width=0.9\textwidth]{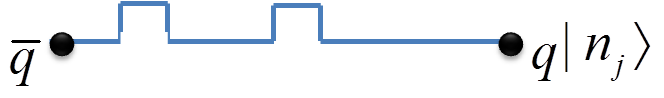}
\caption{A new string state differing by a plaquette action}
\label{p2}
\end{minipage}
\end{figure}

At this point we choose an effective Hamiltonian from the original Kogut-Susskind Hamiltonian for which the interaction part is denoted pictorially in figure \ref{effh}.
\begin{figure}[h]
\label{effh}
\begin{center}
\includegraphics[width=0.8\textwidth,height=0.1\textwidth]{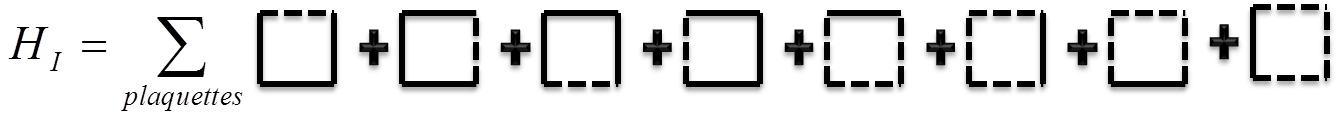}
\end{center}
\caption{Effective interaction Hamiltonian} 
\label{effh}
\end{figure}
Such an effective interaction Hamiltonian, when acts on an arbitrary string state, replaces the links on the string by a staple or vice-versa. Hence, the resultant string state and the original string state do always have a difference in their length by two units and belong to two consecutive excitation level. Such an action is illustrated in figure \ref{p1} and \ref{p2}.
%
Clearly, the state with the minimum length of string, say $l_0$ is the ground states of the system with unperturbed energy $E_0^0=3l_0/4$. The first excited states are of length $l_0+2$ with energy $E_0^1=\frac{3}{4}(l_0+2)$. In general $n^{th}$ excited state has its energy eigenvalue $E_0^n= \frac{3}{4}(l_0+2n)$. The chosen effective interaction Hamiltonian, acting on a particular string state at a particular unperturbed energy level moves it to the next lower or upper energy states respectively. However, as the interaction Hamiltonian is summed over all plaquettes of the lattice, while acting on a string state, it actually create linear combination of many different states, all of which has the same unperturbed energy. Hence, even if the ground state is non-degenerate, the degeneracy increase with increasing energy level\footnote{We will show in the next section, that in reality, this degeneracy factor rises exponentially with excitation.}. 

Let us now concentrate on the matrix element of the interaction Hamiltonian between two string states at two adjacent energy levels. This expression is calculated analytically within prepotential framework. But as the number of string states at each excited level grows rapidly, one needs to incorporate numerical simulations of all such configurations finally yielding a semi analytic treatment of the problem.

As stated earlier, the dynamics in prepotential formulation is calculated locally at each site as given in the expression (\ref{dyna1}). In the perturbation expansion, we have the terms $\langle n_i|H_I|n_j\rangle$. The states $|n_i\rangle$ and $|n_j\rangle$ are identical except around one single plaquette touching the string (an example of such states are shown in figure \ref{p1} and \ref{p2}). Whenever these two states are identical, they will contribute a factor of $1$ in the matrix element. The non-trivial contribution, (around only one plaquette around which these two states actually differ at each step) calculated using (\ref{dyna1}) yields a factor of $1/2$. The dynamics obtained in terms of $18j$ symbol is rotationally symmetric and hence yield the same factor for one plaquette  deformation of the string along any direction. 

Hence we have, for any two string states $|n_i\rangle$ and $|n_j\rangle$,
\bea
\langle n_i|H_I|n_j\rangle &=& \frac{1}{2} ~~~~  \mbox{if the two states differ by only one plaquette} \\
&=& 0  ~~~~ \mbox{ otherwise}
\label{hammat}
\eea

In the next section we prescribe a new numerical simulation technique to simulate all possible loop states of lattice gauge theory. Although, in this paper we will confine ourselves to the above discussed string states connecting static quark-antiquark pair, the general technique applies to simulation of all possible loop configurations of the theory.

\section{Numerical Simulation of Loop States}

\begin{figure}[ht]
\begin{minipage}[b]{.5\textwidth}
\centering
\includegraphics[width=0.6\textwidth]{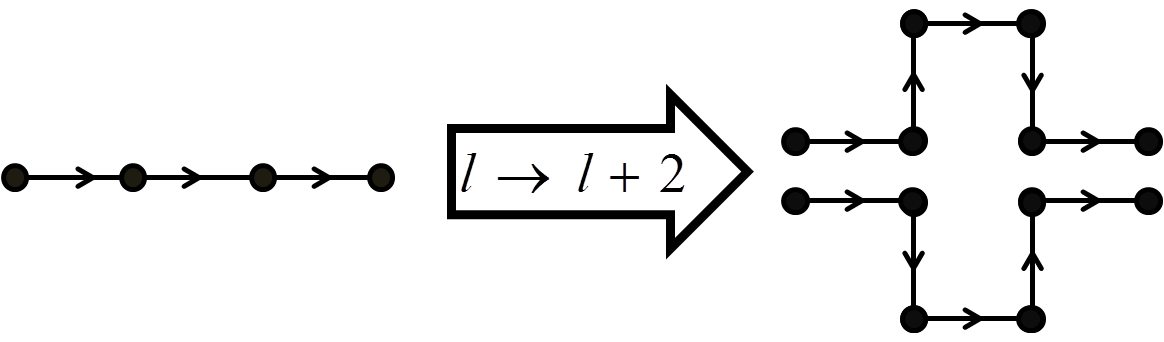}
\caption{BFACF Move: 1}
\label{bfacf1}
\end{minipage}
\hfill
\begin{minipage}[b]{.5\textwidth}
\centering
\vspace{0.1cm}
\includegraphics[width=0.6\textwidth]{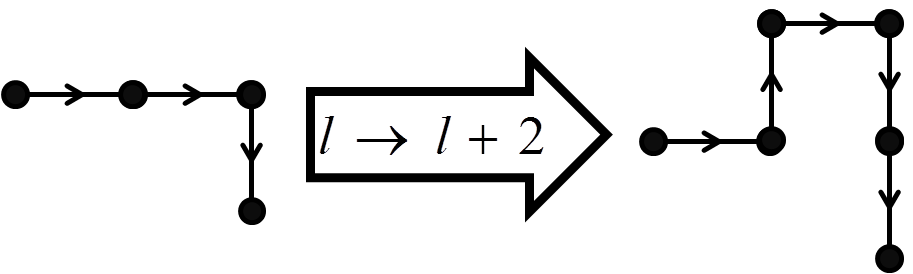}
\caption{BFACF Move: 2}
\label{bfacf2}
\end{minipage}
\end{figure}

\begin{figure}[ht]
\begin{minipage}[b]{.5\textwidth}
\centering
\includegraphics[width=0.6\textwidth]{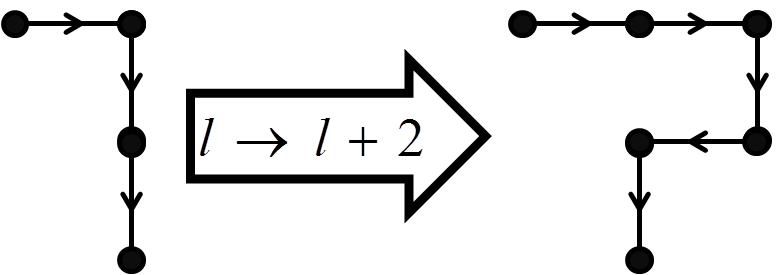}
\caption{BFACF Move: 3}
\label{bfacf3}
\end{minipage}
\hfill
\begin{minipage}[b]{.5\textwidth}
\centering
\vspace{0.1cm}
\includegraphics[width=0.6\textwidth]{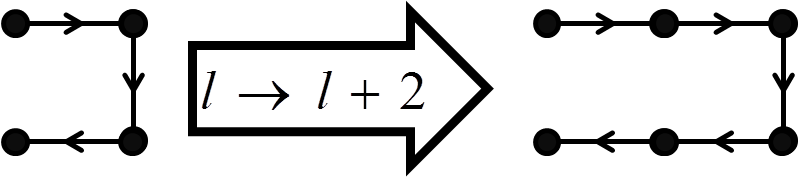}
\caption{BFACF Move: 4}
\label{bfacf4}
\end{minipage}
\end{figure}

Let us now prescribe a new simulation technique for simulating all possible gauge invariant configurations on the lattice starting from the strong coupling vacuum. 
The action of the interaction Hamiltonian on the string (or loop) states can be numerically implemented by BFACF (Berg-Forster-Aragao-Caracciolo-Frohlich) algorithm \cite{bfacf}. The fundamental BFACF moves were originally constructed in context of knotted polygons. We observe that, the entire lattice loop dynamics can indeed be implemented numerically by BFACF moves starting from any arbitrary  loop states. However, in this work, we have chosen a reduced effective Hamiltonian, yielding the dynamics simulated by only the length changing BFACF moves for any loop configuration. Keeping the fixed length BFACF moves as well would have simulated the dynamics of the original Hamiltonian instead of the chosen effective one. Precisely, BFACF algorithm consists of a set of moves, which makes the deviation from one possible path connecting any two points (or even a closed path) on a lattice to another. The path can be considered as a set of links connected with each other as well. Now, the length changing BFACF moves allow one particular link to change to a set of three links as shown in figures \ref{bfacf1}, \ref{bfacf2}, \ref{bfacf3}, \ref{bfacf4}  and the rotations such that the length of that particular path changes by two units. In this algorithm, depending upon the relative orientation of three consecutive links, the middle one gets modified as shown in the figure. Note that, application of these moves on each and every link of a particular loop exactly simulates each and every loop configurations at the next excited level, and hence using this exact algorithm one can produce each and every configurations relevant for the problem. Being an exact simulation technique, it is free from any statistical error as well.

%

\begin{figure}[h]
\label{tree}
\begin{center}
\includegraphics[width=0.6\textwidth,height=0.6\textwidth]{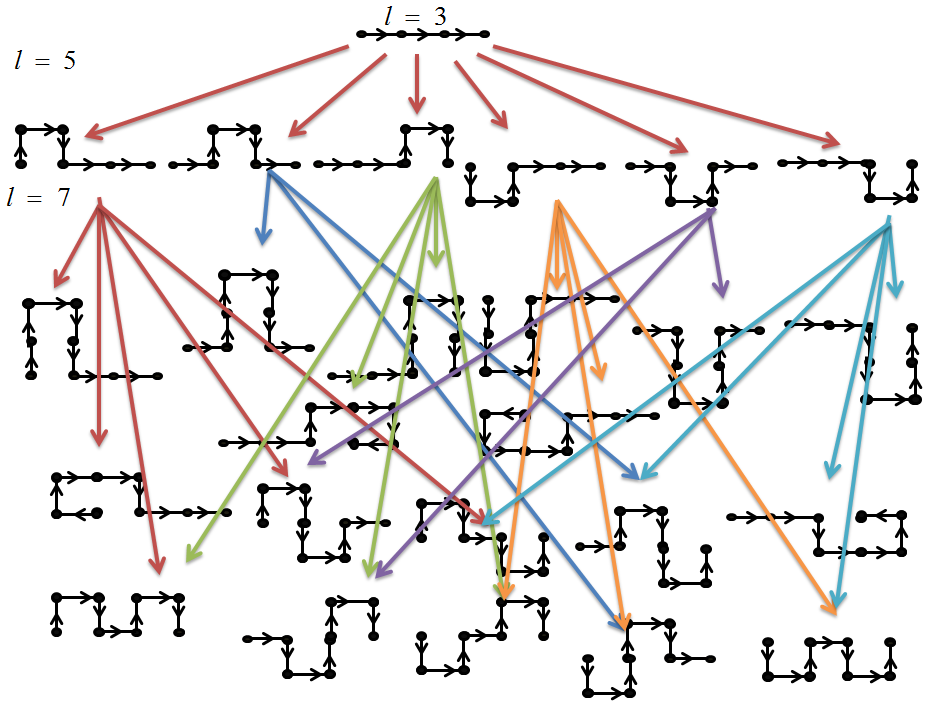}
\end{center}
\caption{Started with a string of length 3 units, all possible strings have been created upto length 7 units. The strings of length 5 units are all created from the initial state. But for the next level of states, which are created from different parents of length 5 units can have multiple parents (denoted by more than one arrow meeting at a particular state). This feature remains valid for higher excited states which have not been illustrated explicitly. Note that, the arrows in the states are not relevant for SU(2) string states, but are important SU(3) and higher rank groups}
\end{figure}

In this work, the configuration of interest is the string state connecting a quark anti-quark pair. We choose our lattice axis in such a way, that both the sources lie on the X-axis. Hence, the shortest distance between them is always a straight line along the axis of length $l_0$, which is the ground state of unperturbed Hamiltonian.
 We now construct all possible higher excited states from this state following the tree structure outlined below. 
\begin{enumerate}
\item Start with an initial state (or, a path) with X coordinates varying from $0,1\ldots, l_0$ and Y coordinate always zero. Length of the path is $l_0$.  Assign a level index $s_1$ and  a serial number (say $p_1^{s_1}$) and. This is now a parent state at level $s_1$.
\item Start from one end of the path $p_1^{s_1}$ and consider each link of the path at a time. For each link, apply the appropriate BFACF move. Check if the move makes the path overlapping or backtracking. If not, the path created is a new path. Assign a particular serial number to each of them.
\item repeat step 2 until one reaches, the other end of the path.
\item In this step, a set of paths  are created  at a level, say $s_2$ with length $l_0+2$ which are say, $q_{s_2}$ in number. All of these paths has the common parent ($p_1^{s_1}$). Assign the serial numbers to  $p_1^{s_2}, p_2^{s_2}\ldots p_{q_{s_2}}^{s_2}$ to each of these paths.
\item Now, each of these paths $p_1^{s_2}, p_2^{s_2}\ldots p_{q_{s_2}}^{s_2}$ will act as parents to create a set of paths in level $s_3$. 
\item Start with path $p_1^{s_2}$, repeat the steps 1, 2, 3 to create all possible paths of length $l_0+4$. These are the paths at $s_3$ with parent $p_1^{s_2}$. 
\item  Then consider path $p_2^{s_2}$ onwards and follow the step 1 and 2 to create a new path at level $s_3$. For each new path created, check with already created paths at the same level, i.e $s_3$, whether this path is indeed a new path or not. If yes, assign a new path number to it and if not, assign  the path, with which it is overlapping, a new parent from which this new path has been created. 
\item When exhausted with all paths at $s_2$, consider all the paths created in level $s_3$  as parent paths to create level $s_4$ following exactly same prescription as in 5, 6, 7.
\item Continue this tree structure to create states depending upon upto which order you want to continue in strong coupling correction
\end{enumerate}
The above algorithm is illustrated pictorially in figure \ref{tree} for a specific example, i.e for $l_0=3$ and upto level $s_3$

The most important feature of the simulation algorithm is as follows:
\begin{itemize}
\item First of all, the above algorithm is an exact simulation algorithm and hence is free from any statistical error.
\item The simulation is in loop space and hence is free from any gauge redundancy.
\item starting from the strong coupling vacuum (or ground state), all possible loop configurations are built at each excited state contributing to correction of strong coupling result. Note that, this scheme involves no further truncation scheme unlike the other loop simulation techniques such as coupled cluster algorithm.
\item The algorithm creates states on an truly infinite lattice.
\end{itemize}

The next part of the calculation involves evaluation of the perturbation expansion taking into account all the states simulated. Actually, for the sake of the present problem it is not necessary to store all the detail of the states as the non-zero matrix element of the interaction Hamiltonian is always the same as given in \ref{hammat}. The matrix element is non-zero only for the two states which are parent-child of each other. In the algorithm described above, we basically construct a parent-children directory for states and keep track of the excitation level of each. Finally, we utilize that to compute the expression given in (\ref{p2}) for perturbation correction order by order. The observations are discussed in the next section.

\section{Results}

First of all, starting with our effective Hamiltonian given in figure \ref{effh}, we create all possible string states, starting from an initial string state of length $l_0$ which is the ground state of the unperturbed Hamiltonian. We tabulate the number of string states produced at each energy level upto next $8^{th}$ excited levels in table \ref{tab1}. 
\begin{figure}[t]
\label{dos}
\begin{center}
\includegraphics[width=0.6\textwidth,height=0.5\textwidth]{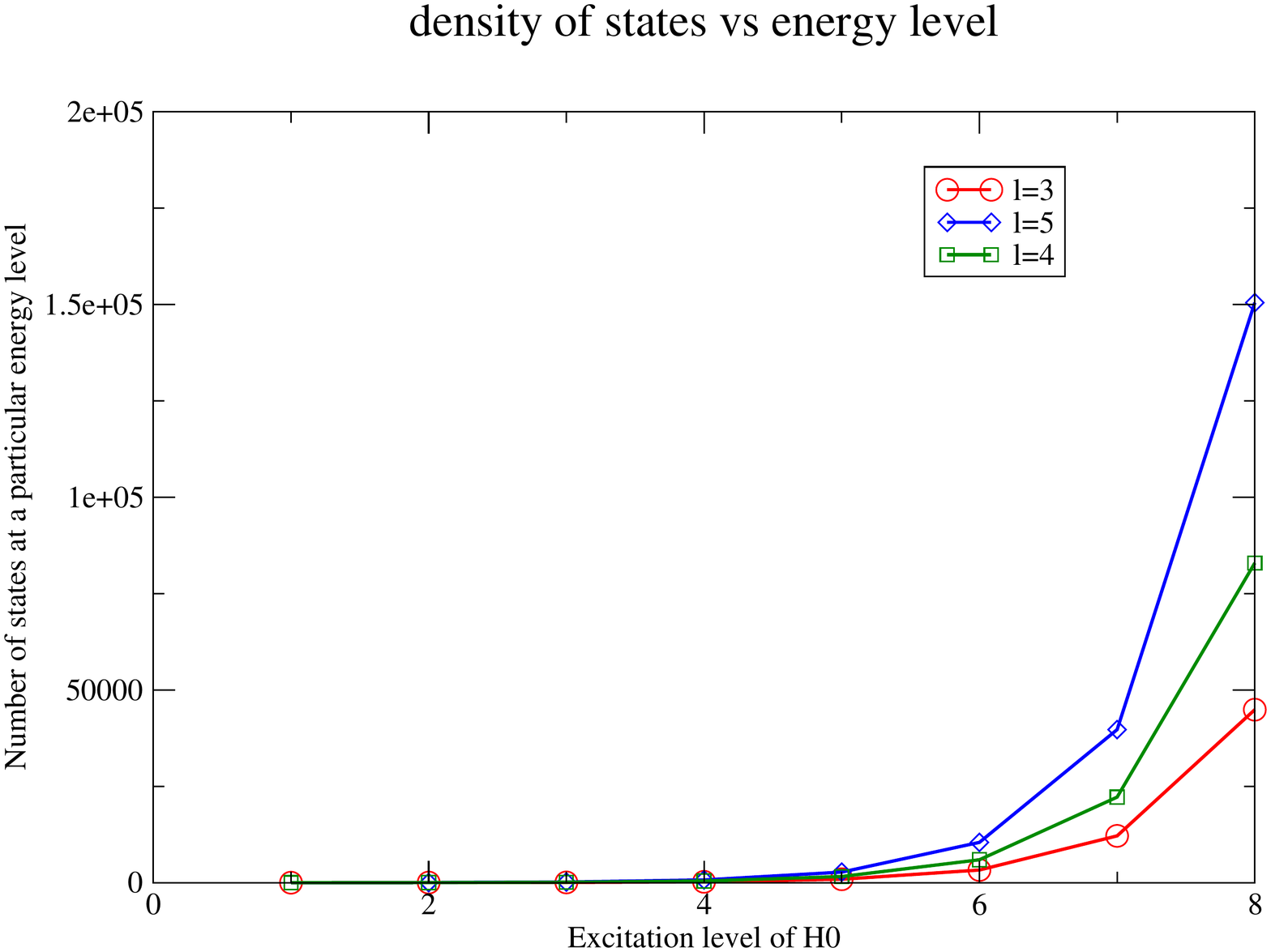}
\end{center}
\caption{We fit the data in table \ref{tab1} in function $N=a+b~e^{cn}$, where, $N$ is the number of states at $n^th$ excited level and $a,b,c$ are fitting parameters the values of which are given in table \ref{tab2}. } 
\end{figure}
\begin{table}[h]
  \begin{center}
    \caption{Degenaracy factor at each excited level of unperturbed Hamiltonian}
    \label{tab1}
    \begin{tabular}{|l|c|c|c|}
    \hline
length of the ground state $(l_0)$ & 3 & 4 & 5\\
      \hline
No. of states with length $l_0$ & 1 & 1&1\\
      \hline
No. of states with length $l_0+ 2$ & 6& 8&10\\
      \hline
No. of states with length $l_0+ 4$ & 18 & 30 &  46\\
      \hline
No. of states with length $l_0+ 6$  & 64 & 108 & 176\\
      \hline
No. of states with length $l_0+8$ & 242 & 416 & 694\\
      \hline
No. of states with length $l_0+10$ & 896 & 1584 & 2720\\
      \hline
No. of states with length $l_0+12$ & 3290 & 5938 & 10448\\
      \hline
No. of states with length $l_0+14$ & 12134 & 22192 & 39732\\
      \hline
No. of states with length $l_0+16$ & 44884 & 82932 & 150474\\
      \hline
    \end{tabular}
  \end{center}
\end{table}
We plot this density of states in figure \ref{dos} and find out that the number of states do grow exponentially, the exponent giving notion of a non-zero Hagedorn temperature present in the theory \cite{Dalley:2004ca} given by the values of the fitting parameter $c$ in the last row of table \ref{tab2} for different string states with different initial length $(l_0=3,4,5)$.

\begin{table}[h]
  \begin{center}
    \caption{Fitting $N=a+b~e^{cn}$, c denoting the Hagedorn temperature}
    \label{tab2}
    \begin{tabular}{|l|c|c|c|}
    \hline
 $l_0$ & 3 & 4 & 5\\
      \hline
a  & 2.76577 & -4.04703 &  -23.98411\\
      \hline
b & 1.28253 & 2.18281 & 3.55591\\
      \hline
c & 1.30787 & 1.31815 &  1.33164\\
      \hline
    \end{tabular}
  \end{center}
\end{table} 

\begin{figure}[h]
\label{evsl}
\begin{center}
\includegraphics[width=0.4\textwidth,height=0.35\textwidth]{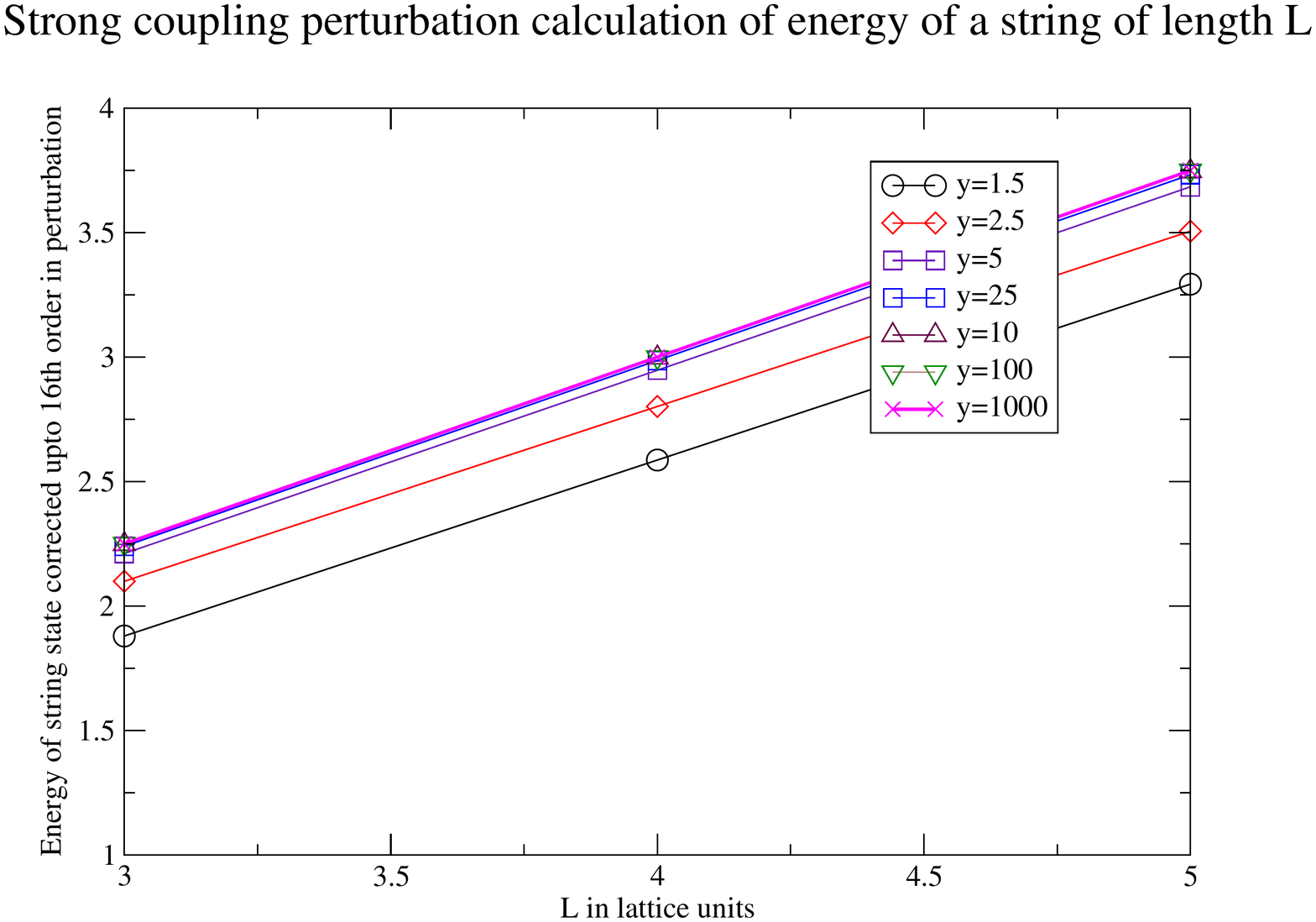}
\end{center}
\caption{ Slope of each graph gives the value of string tension for each case.} 
\end{figure}

\begin{figure}[h]
\label{tvsy}
\begin{center}
\includegraphics[width=0.8\textwidth,height=0.5\textwidth]{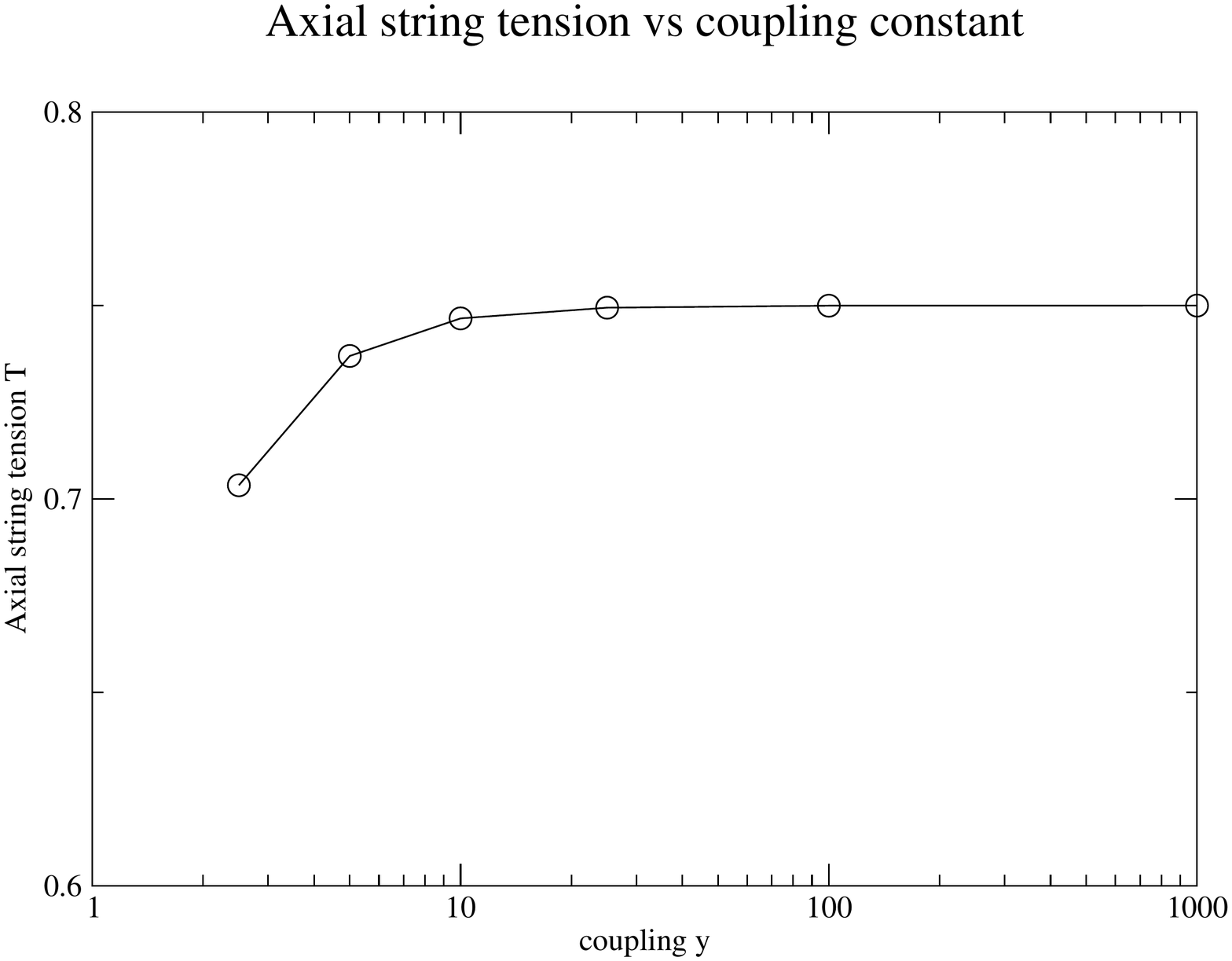}
\end{center}
\caption{We fit the graph to the equation $T=a_0+a_1\ln y$, with $a_0=0.721492$ and $a_1=0.00550585$.} 
\end{figure}

Having generated all the string states connecting a pair of static quark-antiquarks, we calculate the perturbation expansion for the ground state energy of the system following (\ref{Ecorr}) and (\ref{p2}) corrected upto $16^{th}$ order taking care of contributions from all the states simulated (counting total of $61535, 113209, 204301$ for strings with minimum length $3,4,5$ lattice units respectively). In figure \ref{evsl}, we plot the corrected ground state energy of the string state against length of the string state for various values of the coupling $y$, ranging from 1.5 to 1000. For each case we find a linear graph with the slopes denoting constant string tension. 
We further plot the slope of each straight line in figure \ref{evsl} against the coupling for each case in figure \ref{tvsy}. 
Extrapolating the graph to smaller values of the coupling, i.e, beyond the roughening transition point (around $y=1$), we find the following estimate of the string tension as given in table \ref{tab3}.
\begin{table}[h]
  \begin{center}
    \caption{Continuum extrapolation of string tension}
    \label{tab3}
    \begin{tabular}{|l|c|}
    \hline
y & T\\
      \hline
0.001 & 0.688549\\
      \hline
0.0001 & 0.670781\\
      \hline
0.000001 &  0.645426\\
      \hline
    \end{tabular}
  \end{center}
\end{table}

\vspace{5cm}

\newpage

\section{Summary and Discussions}

This work, being the very first attempt exploit the prepotential formulation of lattice gauge theory from numerical perspective, contains an exact simulation algorithm for simulating all possible loop configurations of the theory and some preliminary results illustrating the viability of this new scheme of computations. We use BFACF algorithm to generate the loop configurations of the theory, starting from strong coupling ground state. Note that, each and every possible loop configurations are generated in this algorithm on an infinite lattice. Hence, except the strong coupling limit, no abrupt truncation of loop states is considered. This exact algorithm generates the loops staring from a parent state following a tree structure. The information of the tree structure or the network with all the loops acting as a node enables one to compute strong coupling perturbation expansion using the analytically calculated local dynamics of the theory in prepotential formulation. 

One of the most important feature of prepotential formulation is that, it gives local description of the theory as well as its dynamics. This enables one to split the link operator and hence the full plaquette operator in the Hamiltonian into a set of plaquette operators as given in  figure \ref{h16}. This splitting enables one to construct different effective Hamiltonians of the theory keeping all the symmetries of the system intact. Of course this choice of an effective Hamiltonian depends upon the problem one is interested in. For example, in this work we have chosen an effective  Hamiltonian, which causes the dynamics of the theory only within the connected loops or strings. The disconnected loops do never appear in this modified effective theory. Moreover, our chosen set of plaquette operator acting on any loop state, always created a loop state with length two unit more or less. But there exists some terms in the  Hamiltonian for which one plaquette action deforms a loop in such a way that the length remains same. We have particularly removed those terms from our effective Hamiltonian. Inclusion of such terms would have made the tree to grow sideways as well as opposed to the downward growth as illustrated in figure \ref{tree}. Inclusion of such terms in the algorithm requires modification by including length changing BFACF moves together with appropriate change in calculational scheme. It is indeed important to move forward with this next level of complications by including those terms as well, in order to obtain the dynamics within all connected loops and to compare it with the results obtained in coupled cluster approximation. Moreover different effective Hamiltonians obtained within prepotential framework is worth investigating in order to establish the desired universality between these. Work is these directions are already in progress and will be reported soon.

Finally, this algorithm is applicable beyond lattice gauge theory wherever loops are constructed on a square lattice.

\section*{Acknowledgement}The author would like to thank SS Manna for fruitful discussions regarding implementation of this algorithm and Rudranil Basu for many discussions and help during development of the code. The author would also like to thank Ramesh Anishetty, Manu Mathur and Sanatan Dighal for useful discussions.

\end{document}